\documentclass[mathleft,a4paper
]{an}
\usepackage{graphicx}
\usepackage{times}
\begin{document}


\title{Effective Field Theory for Neutron Stars with Genuine
Many-body Forces}

\author{C.A.Z. Vasconcellos\inst{1}\fnmsep\thanks{Corresponding author:
  \email{cesarzen@cesarzen.com}\newline}  \and R.O. Gomes\inst{1} \and V. Dexheimer\inst{2} \and R.P. Negreiros\inst{3}
\and J. Horvath\inst{4} \and D. Hadjimichef\inst{1}\newline}

\titlerunning{Effective Field Theory for Neutron Stars with Genuine
Many-body Forces}
\authorrunning{C.A.Z. Vasconcellos et al.}

\institute{Instituto de F\'isica, Universidade Federal do Rio
Grande do Sul, Av. Bento Gon\c{c}alves 9500, CEP  91501-970, Porto
Alegre, RS, Brazil \and Department of Physics, Kent State
University, Kent OH, 44242 USA \and Departamento de F\'isica,
Universidade Federal Fluminense, Av. Gal. Milton Tavares de Souza,
s/n�, Campus da Praia Vermelha, CEP 24210-346, Niter\'oi, RJ,
Brazil \and Instituto de Astronomia, Geof\'isica e Ci\^encias
Atmosf\'ericas, Rua do Mat\~ao, 1226, Cidade Universit\'aria, CEP
05508-090, S\~ao Paulo, SP, Brazil}

\received{Dec 2013} \accepted{2014} \publonline{later}

 \keywords{Neutron stars, many-body forces, effective theory}

\abstract{%
The aim of our contribution is to shed some light on open
questions facing the high density nuclear many-body problem. We
focus our attention on the conceptual issue of naturalness and its
role for the baryon-meson coupling for nuclear matter at high
densities. As a guideline for the strengths of the various
couplings the concept of naturalness has been adopted. In order to
encourage possible new directions of research, we discuss relevant
aspects of a relativistic effective theory for nuclear matter with
{\it natural} parametric couplings and genuine many-body forces.
Among other topics, we discuss in this work the connection of this
theory with other known effective Quantum Hadrodynamics (QHD)
models found in  literature and how we can potentially  use our
approach to  describe new physics for neutron stars. We also show
some preliminary results for the equation of state, population
profiles and mass-radius relation for neutron stars assuming local
charge neutrality and beta equilibrium.
 }

\maketitle

\section{Introduction}

Quantum Chromodynamics (QCD) with quark and gluon degrees of
freedom is the underlying theory of the strong interaction
physics. However, at energies relevant for most nuclear phenomena,
hadron and meson fields provide a more appropriate and convenient
description. In this domain, effective quantum field theory (EFT)
and, more precisely, Quantum Hadrodynamics (QHD) represent
efficient frameworks for describing the nuclear many-body problem
as a relativistic system of baryons and mesons.

The motivation for using QHD in this domain can be synthesized in
a few words: simplicity, efficiency, consistency. There are many
arguments that vindicate this assertion: it is the simplest
approach for describing nuclear matter and nuclear systems based
on a local Lorentz-invariant lagrangian density; it represents an
efficient way to parameterize an S matrix or other observables in
nuclear matter; this approach is consistent with analyticity,
unitarity, causality, cluster decomposition, parity conservation,
spin and isospin symmetries, and spontaneously breaking of chiral
symmetry for systems undergoing strong interaction.

Moreover, since the most relevant global phenomena in nuclear
physics are confined to longer length scales, ie, in the
interaction sector where less massive meson fields do\-mi\-na\-te,
 it is not
necessary to explicitly include dynamics at significantly shorter
length scales, where the presence of the most massive meson fields
becomes crucial.  Thus, in describing global properties of nuclear
systems through effec\-ti\-ve field theory, one can ignore the
latter by {\it in\-te\-grating out} heavier degrees of freedom
corresponding to shorter length scales. It is important to notice,
however, the effects of these heavier degrees of freedom are still
implicitly contained in the coupling parameters of the effective
field theory (Serot \& Walecka ~1997). EFT and QHD contain on the
other hand an infinite number of interaction terms and degrees of
freedom. Thus, one needs an {\it organizing principle} to make
sensible calculations.

As a conventional way of classifying and organizing the
interaction terms in our EFT approach as well as a guideline for
the strengths of the various couplings the concept of naturalness
has been adopted.

\subsection{Naturalness}

We take an interaction Lagrangian, which involves the $\sigma$
isoscalar-scalar meson and the $\omega$ isoscalar-vector meson
fields coupled to the nucleon field, defined  in Serot \& Walecka
(1997) as
\begin{eqnarray}
 \!\!\!\!\! {\cal L} =  \sum_{i,k} \frac{c_{i,k}}{i!k!}
(\frac{\sigma}{f_{\pi}})^{i} (\frac{\omega}{f_{\pi}})^{k} \left(
\frac{\partial}{M}\right) (\frac{\bar{\psi}\Gamma
\psi}{f^2_{\pi}M})^{\ell} f_{\pi}^2 \Lambda^{2}  \, ,  \label{1}
\end{eqnarray}
with unknown expansion coefficients $c_{i,k}$ (overall coupling
constants). In the expression above, $\psi$ represents a baryon
field, $\Gamma$ is a Dirac matrix, and derivatives are denoted by
$\partial$. The resulting expressions of the Lagrangian density
also contain factorial counting factors, needed to accomplish
normalization. A direct generalization of this expression may
involve additional fields such as  the $\pi$, $\varrho$ and
$\delta$ meson fields as well as the photon.

There is evidence from studies on properties of ordinary nuclear
matter, that the expansion in the nonlinear meson coupling sector
quickly converges. In fact, keeping only the cubic and quartic
couplings of the $\sigma$ meson one obtains, at saturation and
intermediate densities, a semi-quantitative fit of nuclear matter
properties (Boguta \& Bodmer ~1997; Serot \& Walecka ~1986).
Nevertheless, a controlled expansion of the Lagrangian density to
significantly higher densities ($\rho \geq 5 \rho_0$), as those
found in the interior of neutron stars, requires some assumptions
on a natural ordering of the expansion coefficients. To accomplish
that goal, we may express equation (\ref{1}) in the form
\begin{eqnarray}
 \!\!\!\!\!\! \! \! {\cal L} \! = \!  \left( \frac{\partial
\, or \, m_{\pi}}{M}\right)\!\! (\frac{\bar{\psi}\Gamma
\psi}{f^2_{\pi}M})^{\ell} f_{\pi}^2 \Lambda^{2} \! \sum_{i,k}
\frac{{\tilde{c}}_{i,k}}{i!k!} (\frac{g_{\sigma} \sigma}{M})^{i}
(\frac{g_{\omega} \omega}{M})^{k} \, ,   \label{2}
\end{eqnarray}
where, in this expression, the Goldberger-Treiman relation ($g_s
f_{\pi} \sim M$) was used to eliminate the $f_{\pi}$ factor in the
sum. The assumption of naturalness in the strong interaction
physics means that, unless a more detailed explanation exists, all
conceivable action terms that preserve the required fundamental
symmetries should appear in the effective action of a theory with
natural coupling coefficients  (Georgi ~1993; Manohar ~1996).
Thus, the naturalness condition, when applied to an effective
theory of the strong interaction, establishes that once the
appropriate dimensional scales have been extracted using the {\it
naive dimensional analysis} proposed by Georgi (1993) and Manohar
(1996),  the remaining dimensionless coefficients appearing in the
effective action should all remain of order unity.

If the naturalness assumption is valid, then the effective strong
interaction Lagrangian density can be truncated, with  an
acceptable confidence, within the phenomenological physical domain
of the theory. The  {\it naive dimensional analysis} when applied
in the formulation of an effective Lagrangian density involving
nucleons and strongly interacting meson fields may be synthesized
as follows: the amplitude of each strongly interacting field in
the lagrangian, i.e. the meson fields, becomes dimensionless when
divided by the pion decay weak constant; furthermore, to obtain
the correct dimension ((energy)$^4$) for the Lagrangian density,
an overall normalization scale $f_{\pi}^2 \Lambda^2 \simeq
f_{\pi}^2 M^2$, with $M$ denoting the nucleon mass, has to be
included; finally, for identical meson fields, self-interacting
terms of power $n$ and a symmetrization factor $n!$ (for proper
counting) should be included in the formalism. The overall
dimensionless coefficients, after the dimensional factors and
appropriate counting factors are extracted,  are of order ${\cal
O}(1)$ if naturalness holds. Of course, there is no general proof
of the naturalness property, since no one knows how to derive the
effective strong interaction Lagrangian density from Quantum
Chromodynamics (QCD). Nevertheless,  the validity of naturalness
and {\it naive power counting rules} is supported by
phenomenological studies.

 At least two schemes allow a compact summation of
the Lagrangian density (\ref{2}). For $c_{i,\kappa} = 1$, we have
 \begin{eqnarray}
\!\!\!\!\!\! {\cal L} \! = \! \left(\frac{\partial \, or \,
m_{\pi}}{M}
 \right) \!\! \left( \frac{\bar{\psi} \Gamma \psi}{f_{\pi}^2 M}
\right)^{\ell} \exp \left( \frac{\sigma}{M} + \frac{\omega}{M}
\right)  f_{\pi}^2 \Lambda_{\chi}^2 \, ;
\end{eqnarray}
and for $ c_{i,\kappa} = i! \kappa! $ we obtain
\begin{eqnarray}
\!\!\!\!\!\! {\cal L} = \! \left(\frac{\partial \, or \,
m_{\pi}}{M} \right)  \!\! \left( \frac{\bar{\psi} \Gamma
\psi}{f_{\pi}^2 M} \right)^{\ell} \!\! \left( \frac{1}{\!1 +
\frac{\sigma}{M}} \! \right) \! \left( \! \frac{1}{1 +
\frac{\omega}{M}} \! \right) f_{\pi}^2 \Lambda_{\chi}^2 \, .
\end{eqnarray}

\subsection{Lagrangian Density}

Our strategy in this work is to consider a phenomenological and
more flexible parametrization of the QHD Lagrangian density which
combines the two previous limits and an extension of the
interaction phase space of baryon and meson fields. The complete
expression of our QHD interaction Lagrangian exhausts the whole
fundamental baryon octet ($n$, $p$, $\Sigma^-$, $\Sigma^0$,
$\Sigma^+$, $\Lambda$, $\Xi^-$, $\Xi^0$) and includes many-body
forces simulated by nonlinear self-couplings and meson-meson
interaction terms involving scalar-isoscalar ($\sigma$,
$\sigma^*$), vector-isoscalar ($\omega$, $\phi$), vector-isovector
($\mbox{\boldmath$\varrho$}$) and scalar-isovector
($\mbox{\boldmath$\delta$}$).

The interaction term of our Lagrangian density, ${\cal L}_{int}$,
is defined as
\begin{eqnarray}
& \!\!\!\!\!\!\!\!\!\!\!\! \sum_{B} \!  \bar{\psi}_{B} \! \left[i
\gamma_{\mu}
\partial^{\mu}
  \! - \! \gamma_{\mu}\Sigma^{\mu}_{B \, \xi}\!(\sigma, \sigma^*\!\!\!, \mbox{\boldmath$\delta$}, \omega)
   \! - \!  \gamma_{\mu} \mbox{\boldmath$\tau$} \! \cdot \! \mbox{\boldmath$\Sigma$}_{B
\kappa}^{\mu}\!(\sigma, \sigma^*\!\!\!, \mbox{\boldmath$\delta$},
\mbox{\boldmath$\varrho$}) \right. & \nonumber \\
& \left. \!\!\!\!\!\!\!\!\!\!\!\! \!\!\!\!\!-
\gamma_{\mu}\Sigma_{B \eta }^{\mu}(\sigma, \sigma^*\!\!\!,
\mbox{\boldmath$\delta$}, \phi) \! - \!
 \Sigma^{s}_{B \varsigma}(\sigma, \sigma^*\!\!\!, \mbox{\boldmath$\delta$})  \right] \!  \psi_{B} &
 \end{eqnarray}
where one can identify the following self-energy insertions:
\begin{eqnarray} \!\!\!\!\!\!\!\!\! \Sigma^{\mu}_{B \, \xi} (\sigma, \sigma^*\!\!\!,
\mbox{\boldmath$\delta$}, \omega) & \! = \! &
 g^*_{\omega B \xi}      \omega^{\mu} \, \, ; \,
\mbox{\boldmath$\Sigma$}_{B \kappa }^{\mu}(\sigma, \sigma^*\!\!\!,
\mbox{\boldmath$\delta$}, \mbox{\boldmath$\varrho$})  =
\frac{1}{2} g^*_{\varrho B \kappa }
              \mbox{\boldmath$\varrho$}^\mu \, \nonumber \\
\!\!\!\!\!\!\!\!\! \Sigma_{B \eta }^{\mu}(\sigma, \sigma^*\!\!\!,
\mbox{\boldmath$\delta$}, \phi) & \! - \! &  g^*_{\phi B \eta}
  \phi^\mu \,  \, ; \,
   \Sigma^{s}_{B \varsigma}(\sigma,
\sigma^*\!\!\!, \mbox{\boldmath$\delta$}) = M   \Sigma^{s}_{B
\varsigma}  \, .
\end{eqnarray}
In these expressions, $g^*_{\Phi}$, with $\Phi =\sigma, \sigma^*,
\omega,  \phi, \varrho, \delta$, re\-pre\-sents the effective
parametrized baryon-meson coupling constants, defined as
 $g_{\omega B
\xi}^* =  g_{\omega B} m^*_{B \xi}$, $g_{\varrho B \kappa}^* =
g_{\varrho B} m^*_{B \kappa}$, $g_{\phi B \eta}^*  = g_{\phi B}
m^*_{B \eta}$, and  $\Sigma^{s}_{B \varsigma} = m^*_{B
\varsigma}$, with
\begin{equation}
m^*_{B \alpha} \equiv  \left( 1 + \frac{g_{\sigma B} \sigma +
g_{\sigma^* B} \sigma^*+ \frac{1}{2}g_{\delta B}
\mbox{\boldmath$\tau$} \cdot \mbox{\boldmath$\delta$} }{\alpha
M_B} \right)^{-\alpha} \, ;
\end{equation}
($\alpha = \xi$, $\kappa$, $\eta$, $\varsigma$) (Taurines et al.
2001; Vasconcellos et al. 2012). Properties of the fields
considered in our formulation are presented in table
(\ref{campos})).
\begin{center}
\begin{table}[t]
\vspace{-0.55cm}
 \small \caption{Properties of the fields
considered in the formulation (\ref{NLD}). In what follows, we use
the abbreviations: ISS: isoscalar-scalar; IVS: isovector-scalar;
ISV: isoscalar-vector; IVV: isovector-vector.}
\begin{center}
\begin{tabular}{llclc}
\hline
Fields & Classification & Particles & Coupling   & Mass \\
 & & & Constants & (MeV) \\
 \hline
$\psi_B$ & Baryons & N, \,$\Lambda$,& N/A & 939, 1116, \\
         &           &              $\Sigma$, \,$\Xi$   &     &       1193, 1318\\
$\psi_l$ & Leptons & $e^-$, $\mu^-$& N/A &0,5,\,106 \\
\hline
$\sigma $ & ISS-meson  & $\sigma $ & $g^*_{\sigma_B}$ & 550 \\
$\mbox{\boldmath$\delta$}$ & IVS-meson  & $a_0$ &$g^*_{\delta_B}$ &980\\
$\omega_\mu $ & ISV-meson  & $\omega $ &$g^*_{\omega_B}$ &782 \\
$\mbox{\boldmath$\varrho$}_{\mu}$ & IVV-meson  & $\rho$ & $g^*_{\varrho_B}$&770 \\
$\sigma^\ast $ & ISS-meson  & $f_0$ &$g^*_{\sigma\ast_B}$ & 975 \\
$\phi_\mu $ & ISV-meson & $\phi $ &$g^*_{\phi_B}$ & 1020 \\
\hline
\end{tabular}
\label{campos}
\end{center}
\end{table}
\end{center}

 The complete expression for the Lagrangian density in the mean field
 approximation,
 $ {\cal L}_{\xi \varsigma  \kappa \eta }$,
 becomes
\begin{eqnarray}
\!\!\!\!\!\!\!\!\!\!\!\!  & & \frac{1}{2}m_\sigma^2\sigma_0^2 +
\frac{1}{2} m_{\sigma\ast}^2 \sigma_0^{\ast 2}
+\frac{1}{2}m_\omega^2\omega_0^2 +\frac{1}{2} m_\phi^2 \phi_0^2
\\ & \! + \! & \frac{1}{2} m_\rho^2 \varrho_{03}^2 + \frac{1}{2}
m_\delta^2 \delta_{3}^2   + \sum_l \bar\psi_l \,(i \gamma_\mu
\partial^\mu - m_l)\,\psi_l \nonumber
\\  & \! + \! & \sum_B \bar{\psi}_B (i \gamma_\mu \partial^\mu
 -  g_{\omega B} m^*_{B \xi}\gamma_0 \omega_0   -   M_{B \varsigma}^{\ast}  ) \psi_B \nonumber \\  &  \! - \! &
 \sum_B  \!
 \bar{\psi}_B  ( \frac{1}{2} g_{\varrho B} m^*_{B  \kappa}
 \gamma_0
\tau^{(3)} \varrho_{03}   \! + \! g_{\phi B} m^*_{B  \eta} \!
\gamma_0  \phi_0 ) \psi_B
 \, , \nonumber\label{NLD}
\end{eqnarray}
with the effective baryon mass defined as $ M_{B \varsigma}^{\ast}
=  M_B m^*_{B \varsigma}$.

The resulting expression for the Lagrangian density is known as
the parameterized coupling model. Among the numerous possibilities
of choice of parameters for the model we may focus our attention
in variations of the $\varsigma$ pa\-ra\-me\-ter keeping $\xi =
\kappa = \eta = 0$ (scalar or S-model).  This parametrization
reduces to the Walecka's QHD-I model (Serot \& Walecka 1997) if
$\varsigma = 0$ and $g_{\varrho B} = g_{\phi B} = 0$. In case only
$g_{\varrho B} \neq 0$, this parametrization reduces to the
Walecka's QHD-II model (Serot \& Walecka 1997). Assuming $
\varsigma = 1$, performing a binomial expansion of $m_{B
\varsigma}$, truncating the perturbative series to cubic and
quartic self-interactions terms involving the scalar-isoscalar
$\sigma$ meson (making $g_{\sigma^* B} = g_{\delta B} = 0$), this
parametrization reduces to the model of Boguta \& Bodmer (1997).
The second choice contemplates variations of the parameters
$\varsigma$ and $\xi$ while keeping $\kappa = \eta = 0$
(scalar-isoscalar-vector or SISV-model). In the third choice we
may consider variations of $\varsigma$, $\xi$, and $\kappa$, and
fixing $\eta = 0$ (scalar-isoscalar-vector-isovector-vector or
SISVIVV-I-model), as already discussed in (Dexheimer, Vasconcellos
\& Bodmann ~2008). Finally, in the fourth choice we may consider
variations of the four parameters of the theory, $\varsigma$,
$\xi$, $\kappa$, and $\varsigma$
(scalar-isoscalar-vector-isovector-vector  or SISVIVV-II-model).
 These examples of parameterizations of our approach are shown in
 Table
(\ref{parameterizations}).
\begin{center}
\begin{table}[htdp]
\caption{Examples of parameterizations of our model. S: scalar
model; SISV: scalar-isoscalar-vector model; SISVIVV:
scalar-isoscalar-vector-isovector-vector  model. Model II differs
from I, due to the presence of the $\phi$ meson.}
\centerline{\begin{tabular}{ccccc} \hline Model & $\varsigma$ &
$\xi$ & $\kappa$ & $\eta$ \\ \hline
S & $ \neq 0$ & $0$ & $0$ & $0$ \\
\hline SISV & $\neq 0$ & $\neq 0$ & $0$ & $0$ \\
\hline SISVIVV-I & $\neq 0$ & $\neq 0$ & $\neq 0$  & $0 $
\\ \hline SISVIVV-II & $\neq 0$ & $\neq 0$ &$\neq 0$ &$\neq 0$
\\
\hline
\end{tabular} }
\label{parameterizations}
\end{table}
\end{center}

We consider, as an example, the expansion of the term $g_{\omega
B} m^*_{B \xi}\gamma_0 \omega_0$ in expression (\ref{NLD}), in the
particular case $\xi=1$. In the framework of a local mean field
approximation, expectation values of meson fields correspond to
classical numbers\footnote{It is well known that in the realm of
the mean field approximation, expectation values of meson fields
may be treated as classical numbers in space-time despite the
density dependence of these fields in Fermi space.}. In this
context, taking into account that $\frac{g_{\sigma B} \sigma_0 +
g_{\sigma^* B} \sigma^*_0+ \frac{1}{2}g_{\delta B} \tau_3
\delta_{30} }{M_B} << 1$,
 we may use the binomial theorem
to expand the contribution $g_{\omega B} m^*_{B \xi}\gamma_0
\omega_0$ for the particular choice $\xi = 1$:
\begin{eqnarray}
& g_{\omega B} m^*_{B 1}\gamma^0 \omega_0   \sim  g_{\omega
B}\gamma^0 \omega_0 & \nonumber \\ & -  g_{\omega B} \left(
\frac{g_{\sigma B} \sigma_0 + g_{\sigma^* B} \sigma^*_0+
\frac{1}{2}g_{\delta B} \tau_3 \delta_{30} }{M_B}\right)\gamma^0
\omega_0 & \nonumber \\ & +  g_{\omega B} \left( \frac{g_{\sigma
B} \sigma_0 + g_{\sigma^* B} \sigma^*_0+ \frac{1}{2}g_{\delta B}
\tau_3 \delta_{30} }{M_B}\right)^2 \gamma^0 \omega_0 & \nonumber \\
& -   g_{\omega B} \left( \frac{g_{\sigma B} \sigma_0 +
g_{\sigma^* B} \sigma^*_0+ \frac{1}{2}g_{\delta B} \tau_3
\delta_{30} }{M_B}\right)^3 \gamma^0 \omega_0  ... &
\end{eqnarray}
From this expression we can identify many-body attractive and/or
repulsive coupling terms like for instance $\sigma_0 \omega_0$,
$\sigma_0^2 \omega_0$, $\sigma^{*2}_0 \omega_0$, $\delta_{30}
\omega_0$, $\delta_{30}^2 \omega_0$, $\sigma_0  \sigma^{*}
\delta_{30} \omega_0$ and many others. Similarly, the remaining
terms of the Lagrangian density also exhibit additional many-body
coupling terms. Again, the final selection of the contributions to
be considered requires an analysis of the formal coherence of the
theory. In this sense, it is important to remember that the theory
must embody fundamental symmetries, conservation laws as well as
physical properties which are relevant for strong in\-te\-racting
relativistic nuclear many-body systems, such as Lorentz
covariance, microscopic causality, naturalness, analyticity,
uniqueness, among others.


From expression (\ref{1}), an extension of the model of Serot \&
Walecka (1997), which exhausts the whole fundamental baryon octet
($n$, $p$, $\Sigma^-$, $\Sigma^0$, $\Sigma^+$, $\Lambda$, $\Xi^-$,
$\Xi^0$) and includes baryon-meson interaction terms involving
scalar-isoscalar ($\sigma$, $\sigma^*$), vector-isoscalar
($\omega$, $\phi$), vector-isovector ($\varrho$) and
scalar-isovector ($\delta$) may be synthesized as
\begin{eqnarray}
\!\!\!\!\!\!\!\!\!\!\!\!\! {\cal L} & \!\! = \!\! & \sum_B
\sum_{\wp} \! \frac{c_{\wp}}{\Pi_{\wp} \wp! } \left(
\frac{\partial \, or \, m_{\pi}}{M_B}\right) (\frac{\bar{\psi}_B
\Gamma \psi_B}{f^2_{\pi}M_B})^{\ell} f_{\pi}^2 \Lambda^{2}
 \nonumber \\
& \times & (\frac{\sigma}{f_{\pi}})^{i}
(\frac{\sigma^*}{f_{\pi}})^{j} (\frac{\omega}{f_{\pi}})^{k}
 (\frac{\varrho}{f_{\pi}})^{m}
(\frac{\delta}{f_{\pi}})^{n} (\frac{\phi}{f_{\pi}})^{q} \, ,
\label{extension}
\end{eqnarray}
with $\wp = i, j, k, m, n, q$. The {\it natural} limit of this
expression, namely, by assuming $c(\wp) \to 1$, is
\begin{eqnarray}
 \!\!\!\!\!\!\!\!\!\!  {\cal L}   =  \sum_B \left(\frac{\partial
\, or \, m_{\pi}}{M_B} \right) \!\! (\frac{\bar{\psi}_B \Gamma
\psi_B}{f^2_{\pi}M_B})^{\ell}  f_{\pi}^2 \Lambda^{2} \exp
\sum_{\iota}\frac{g_{\iota} \Phi_{\iota}}{M_B} \, ,
\end{eqnarray}
with $\Phi_{\iota}=\sigma, \sigma^*, \omega,  \phi, \varrho,
\delta$. Through systematic choices for the parameters of the
Lagrangian density (\ref{NLD}) we also obtain, in the limit of
naturalness, exponential couplings.

\section{Scalar Model}

In the following, we consider a simplified version of the model
described above. In this version, we analyze the effects of the
long range many-body components of the nuclear force on the
structure of the star. We seek in this way to understand how these
nonlinear contributions affect global properties of neutron stars,
as for instance the equation of state, the stellar mass and the
stellar population of particles in the medium. The understanding
of these effects is important, since the long-range components of
the strong interaction play a fundamental role in the stability of
the star. In a later contribution, we will specifically examine
the corresponding contribution of the short-range components of
the strong interaction, which also play a fundamental role in the
stability of the star. From the global point of view, the
long-range attractive components contribute to stellar
contraction. The short-range components of the strong interaction
act in the opposite direction.

We consider a mean field scalar version of our model with the
explicit inclusion of an Yukawa attractive term in the Lagrangian
density:
\begin{eqnarray}
\!\!\!\!\!\!\!\! {\cal L}_{\lambda} & = &
\frac{1}{2}m_\sigma^2\sigma_0^2 + \frac{1}{2}m_\omega^2\omega_0^2
+  \frac{1}{2} m_\rho^2 \varrho_{03}^2
\\ & + &  \sum_l \bar\psi_l \,(i \gamma_\mu
\partial^\mu - m_l)\,\psi_l \nonumber
\\  & + & \sum_B \bar{\psi}_B (i \gamma_\mu \partial^\mu
 -  g_{\omega B} \gamma^0 \omega_0   -   M_{B \lambda}^{\ast}  ) \psi_B \nonumber \\  &  - &
 \sum_B  \!
 \bar{\psi}_B  ( \frac{1}{2} g_{\varrho B}
 \gamma^0 \!
\tau^{(3)} \! \varrho_{03}) \psi_B
 \, , \nonumber
\end{eqnarray}
where the subscripts $B$ and $l$ label, respectively, the baryon
octet ($n$, $p$, $\Lambda^0$, $\Sigma^-$, $\Sigma^0$, $\Sigma^+$,
$\Xi^-$, $\Xi^0$) and lepton ($e^-$, $\mu^-$) species. For more
details see (Vasconcellos et al. 2012).

Evidently, our model must be in agreement with the experimental
data related to the saturation properties of nuclear matter. We
adopt for the saturation density of nuclear matter $\rho_0= 0.17$
fm$^{-3}$ and for the binding energy $\epsilon_B = -16.0$ MeV. The
$\lambda$ parameter is constrained to describe the nucleon
effective mass at saturation between $0.70-0.78$ MeV. The
isovector coupling constant $g_{\varrho}$, on the other hand, is
chosen to describe the symmetry energy coefficient $a_{asym}=32.5$
MeV; for more details see for instance Haensel, Potekhin \&
Yakovlev (2007). Furthermore, we use the equilibrium properties of
nuclear matter to introduce a coherent set of NM coupling
constants in the saturation density range, $g_{\sigma N}$,
$g_{\omega N}$, $g_{\varrho N}$ (see table (\ref{cc}). In the high
density regime, the HM coupling constants have been fitted to
hypernuclear properties (Schaffner et al. ~1994), with the scalar
coupling constants fixed to the potential depth of the
corresponding hyperon species: $U_{\Lambda} = U_{\Sigma} = - 30$
MeV and $U_{\Xi} = 30$ MeV.

\begin{figure}[h,t,b]
\vspace{-0.2in}
 \centering
 \includegraphics[width=8.0 cm]{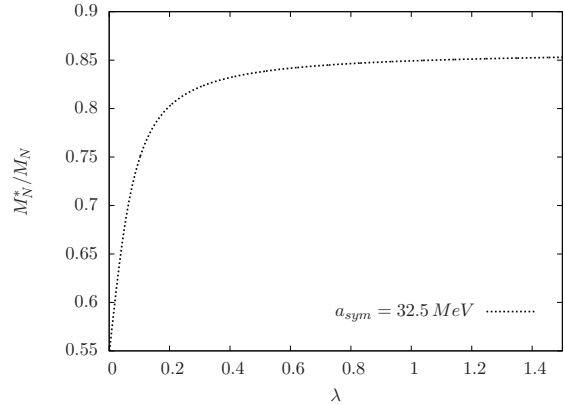}
 \caption{Dependence of the effective nucleon mass at nuclear sa\-tu\-ra\-tion on the $\lambda$ parameter.
 \label{meff}}
 \end{figure}

The effective parameterized nucleon mass  $M^*_{B \lambda}$, for $
\frac{g_{\sigma B} \sigma_0}{\lambda M_B} <<1 $, is defined as
\begin{eqnarray}
\!\!\!\!\!\!\!\!\! M^*_{B \lambda}  \! \simeq  \! M_B \! \left(
\!\! 1 \! - \! \frac{g_{\sigma B}\sigma_0}{M_B} \! + \!\!  \left(
\!
\begin{array}{c} \lambda \\ 2  \end{array} \right) \!\! \left( \! \frac{g_{\sigma B} \sigma_0}{\lambda M_B}  \right)^2  \! \right) \! + \! {\cal O}(3)
,   \label{effectivemass}
\end{eqnarray}
 with ${\tiny
(\begin{array}{c} \lambda \\ 2  \end{array} ) }$ representing the
generalized binomial coefficients of the expansion
 and
where we emphasize the direct dependence of the effective baryon
mass on the $\lambda$ parameter of the model. The dependence of
the effective nucleon mass on the $\lambda$ parameter at nuclear
saturation is shown in Fig. (\ref{meff}).

\begin{figure}[h,t,b]
\vspace{-0.20cm}
 \begin{center}
 \includegraphics[width=8.0 cm]{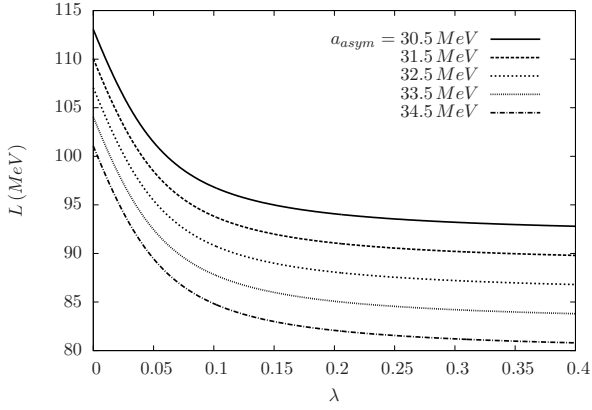} \label{L}
 \caption{Dependence of the slope parameter $L$ on $\lambda$ for
different values of the symmetry energy $a_{\mathrm{asym}}$.}
\label{Llambda}
\end{center}
 \end{figure}

The density dependence of the nuclear matter symmetry energy
($E_{\mathrm{sym}}$) can be characterized in terms of a few bulk
parameters by expanding it in a Taylor series around the nuclear
saturation density $\rho_0$
\begin{eqnarray}
E_{sym}(\rho) & = & E_{sym}(\rho_0)  +  L \! \left( \frac{\rho -
\rho_0}{3 \rho_0}\right)   \nonumber \\ & + & \frac{K_{sym}}{2}
\left( \frac{\rho - \rho_0}{3 \rho_0}\right)^2 +  {\cal O}(3) +
...
\end{eqnarray}
where $E_{\mathrm{sym}}(\rho_0) = a_{\mathrm{asym}}$ is the value
of the symmetry nuclear energy at saturation. In this expression,
$L$ is the slope parameter and the curvature parameter
$K_{\mathrm{sym}}$ is the isovector correction to the compression
modulus, defined respectively as
 \begin{eqnarray}
\!\! \!\!\!\!\!\!\! L   =  3\rho_0 \left(\frac{\partial
a_{\mathrm{asym}}}{\partial\rho}\right)_{\rho =
 \rho_{0}} \!\!\! \!\!\!; \,
 K_{\mathrm{sym}}   =  9\rho_0^2 \left(\frac{\partial^2 a_{asym}}{\partial\rho^2}\right)_{\rho =
 \rho_{0}} \!\!\!\!\!\!\! .
\end{eqnarray}
 Fig. (\ref{Llambda}) shows the dependence of
$L$ on $\lambda$ and on the symmetry coefficient $a_{asym}$. The
slope parameter $L$ is related to $P_0$, the pressure from the
symmetry energy for pure neutron matter at saturation density. The
symmetry pressure $P_0$ provides the dominant baryon contribution
to the pressure in neutron stars at saturation density.

\section{Results and Conclusions}

\begin{figure}[h,t,b]
\vspace{-0.4cm}
 \centering
 \includegraphics[width=8.0cm]{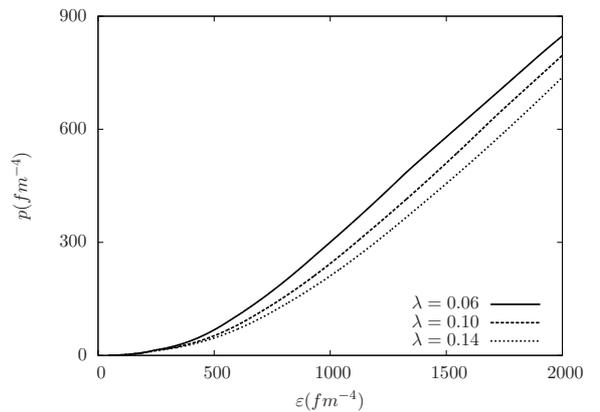}
 \caption{\label{eos} Dependence of the equation of state (EoS) on $\lambda$.}
 \end{figure}

\begin{figure}[h,t,b]
 \centering
 \includegraphics[width=8.0cm]{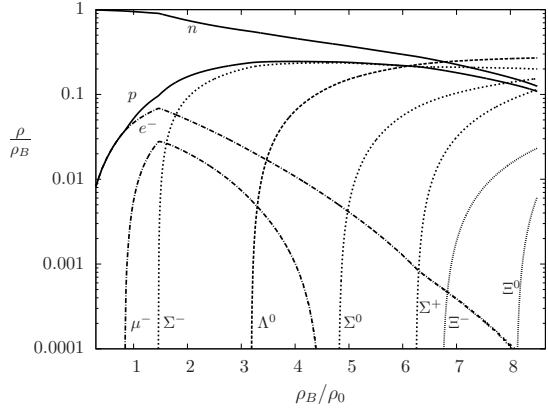}
 \caption{Particle population for $\lambda=0.06$. In the figure,
the labels $\rho_B$ and $\rho_0$ indicate respectively baryon
density and nuclear saturation density.  \label{popl006}}
 \end{figure}

\begin{figure}[h,t,b]
 \vspace{-0.4cm}
 \centering
 \includegraphics[width=8.0cm]{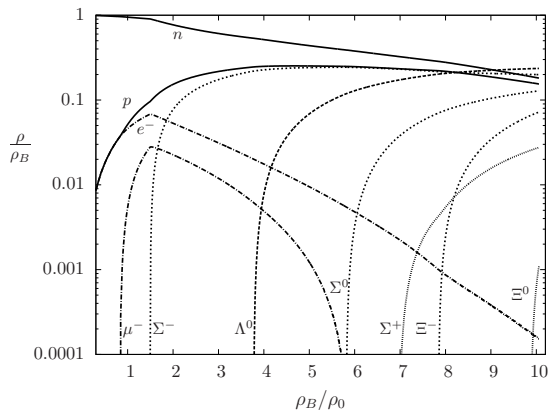}
 \caption{Particle population for $\lambda=0.10$. The symbols have the same meaning as in the previous figure.\label{popl010}}
 \end{figure}

\begin{figure}[h,t,]
 \centering
 \includegraphics[width=8.0cm]{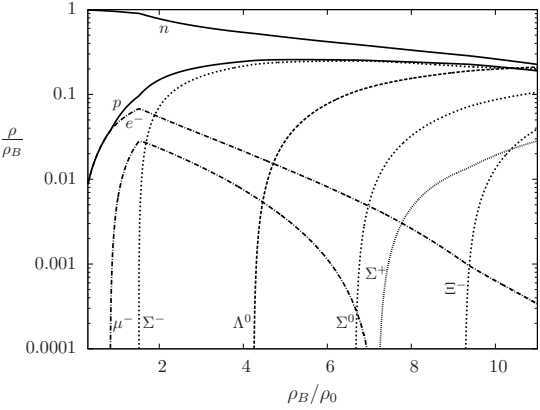}
 \caption{Particle population for $\lambda=0.14$.
Symbols have the same meaning as in previous figures.
 \label{popl014}}
 \end{figure}

\begin{figure}[h,t]
\vspace{-0.4cm}
 \centering
 \includegraphics[width=8.0cm]{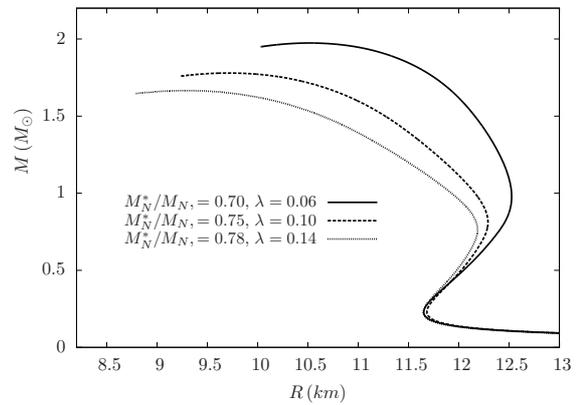}
 \caption{Mass-radius diagram for three different parameterizations
of the scalar version of the model: $\lambda =
0.06,\,0.10,\,0.14$.
 \label{tov}}
 \end{figure}

In the following, we determine the EoS and population profiles for
neutron stars assuming local charge neutrality and beta
equilibrium for the scalar version of our approach. By solving the
Tolman-Oppenheimer-Volkoff (TOV) equations (Tolman 1939;
Oppenheimer \& Volkoff ~1939) we obtain the mass-radius relation
for families of neutron stars with hyperon content.
 Our results for different values of the $\lambda$
parameter are illustrated in Figs. (\ref{eos}), (\ref{popl006}),
(\ref{popl010}), (\ref{popl014}), and (\ref{tov}).
\begin{table}[h,t,b]
\vspace{-0.4cm}
  \caption{\label{cc} Coupling parameters of our approach.}
\begin{tabular}{cccccc} \small
$\lambda$ & $(\frac{M^*}{M})_{_0}$ &  $K_0$~(MeV) &
$(\frac{g_{\sigma B}}{m_{\sigma}})^2$ & $(\frac{g_{\omega
B}}{m_{\omega}})^2$ &
$(\frac{g_{\varrho B}}{m_{\varrho}})^2$ \\
  \hline
    & &  & & & \\
  0.06 & 0.70  & 262   &  11.87  & 6.49  & 3.69  \\
  0.10 & 0.75  & 226   &  10.42  & 5.10  & 3.94  \\
  0.14 & 0.78  & 216   &  9.51   & 4.32   & 4.07 \\
\end{tabular}
\end{table}
In the figures, each value of $\lambda$  generates a sequence of
neutrons stars, where each star will have a different EoS,
particle population, central density, and maximum mass. The
parametrization that generates the higher pressure and maximum
mass of a neutron star is the one with the lower possible value of
the parameter $\lambda$, ie $\lambda = 0.06$, which provides a
mass of $1.97\,M_{\odot}$, in very good agreement with recent
observations (Demorest et al. ~2010). This result corresponds to
the parametrization which provides a smaller effective baryon mass
and, hence, a more bound matter. Although smaller lambda values
originate higher values for the maximum mass of neutron stars,
this parameter should attain a minimum value that allows the model
to reproduce nuclear saturation properties, like for example a
compressibility modulus smaller than 300 MeV.

 The analysis of these
results demands first to remember that a stiffer, or equivalently,
more rigid equation of state of nuclear matter is related to
higher values of the internal pressure of the system and,
accordingly, to higher values of the compressibility modulus
$|K_{sym}|$ of nuclear matter. This in turn requires stronger
contributions from repulsive components of the nuclear force when
compared to the attractive ones. In our general approach, however,
many body forces lowers the intensities both of attractive and
repulsive interaction terms due to {\it shielding effects}, which
result in higher (lower) values of the compressibility modulus
$|K_{sym}|$ of nuclear matter in the case of higher (lower)
relative reduction of the attractive (repulsive) contributions.

Our present results (scalar model) are consistent with the
analysis above, since we have obtained a stiffer equation of state
for lower values of the $\lambda$ parameter ($\lambda = 0.06$),
which corresponds to the stronger relative reduction of the
attractive contribution (see Eq. (\ref{effectivemass})).
Si\-mi\-lar\-ly, the results for the particle po\-pu\-la\-tion
show that the threshold equation for a given species (Glendenning
~1996),
$$\mu_n \! - \! q_B \mu_e \! - \! g_{\omega B}\omega_0 \! - \!
g_{\varrho B} \varrho_{0 3} I_{3 B} \geq M_B ( 1 \! - \! g_{\sigma
B} \sigma_0/M_B),$$ is also affected by the presence of many-body
forces components in the strong interaction sector (sse Eq.
(\ref{effectivemass}):
\begin{eqnarray}
& \!\!\!\!\!\!\!\!\!\!\! \mu_n  \! - \! q_B \mu_e  \! - \!
g_{\omega B}\omega_0 \! - \! g_{\varrho B} \varrho_{0 3} \! I_{3
B} \! \geq \!
  M_B ( 1 \! - \! g^*_{\sigma B} \sigma_0/M_B) & \nonumber \\
& \!\!\!\!\!\!\!\!\!\!\!\!\!\!\!\!  \simeq  \! M_B \! \left( \! 1
- \frac{g_{\sigma B}\sigma_0}{M_B} + \left( \!
\begin{array}{c} \lambda \\ 2  \end{array} \right) \! \left( \!
\frac{g_{\sigma B} \sigma_0}{\lambda M_B} \! \right)^2   \right) +
{\cal O}(3) \, , & \label{TE}
\end{eqnarray}
where $\mu_n$ and $\mu_e$ represent respectively the neutron and
electron chemical potential, and $q_B$ is the baryon charge.
According to Eq. (16) many body forces shift the critical density
for hyperon sa\-tu\-ra\-tion to higher densities and originate an
anti-correlation between the amount of hyperons and the values of
$\lambda$: to the lowest va\-lues of $\lambda$ cor\-res\-pond
larger amounts of hyperons in the system, which may cause the
softening of the EoS. However, our results show that the shielding
effect of the long-range nuclear components of the nuclear force
is predominant. The first hyperon species that appears is the
$\Sigma^{-}$, closely followed by the $\Lambda$, since the
negative charge of the $\Sigma^{-}$ outweighs the 80 MeV mass
difference, as a result of the more lenient conditions
$\mu_{\Lambda} = \mu_n$ and $\mu_{\Sigma^-} = \mu_n + \mu_e$. The
$\Sigma^{-}$ fraction saturates at about $0.3$, while $\Lambda$,
free of isospin-dependent forces, continue to accumulate until
short-range repulsion forces cause them to saturate. Other hyperon
species follow at higher densities.


It is our understanding that the naturalness condition is
equivalent to exhaust the phase space of the fundamental
interactions. This depletion of the phase space can be
accomplished by including in nuclear matter the largest possible
number of {\it information}, ie. baryon and meson degrees of
freedom and additionally considering many body forces between
baryon and meson fields, as well as self-coupling terms involving
meson fields. When the condition of na\-tu\-ral\-ness is achieved,
it is unnecessary to get rid of hea\-vier degrees of freedom by
integrating then out. In this case, the effects of heavier degrees
of freedom are not anymore implicitly contained in coupling
parameters of the effective field theory. An interesting aspect of
our study is that appa\-ren\-tly, many body forces occupy a larger
role than originally thought in the description of properties of
nuclear system at high density.
Moreover,
the condition of naturalness is achieved with only a small part of
the parameter set. This result was expected and can be explained
by the saturation property of nuclear matter: for parameter values
less than $3$, the model with parameterized couplings completely
exhausts the phase space of many-body interactions.

The model with parameterized couplings shows promising results.
However, the model needs to broaden its scope by the adoption of
new parameterizations, expanding the set of parameters of the
theory. Interesting issues for future studies will also be the
role of finite temperature, neutrino trapping and strong magnetic
effects in neutron stars. Work along these lines is in progress.

\end{document}